\documentclass[11pt, preprint]{aastex}

\shorttitle{H$_2$ Fluorescence in Eridanus} \shortauthors{Ryu et
al.}

\begin{document}


\title{Molecular Hydrogen Fluorescence in the Eridanus Superbubble}


\author{K. Ryu\altaffilmark{1}, K. W. Min\altaffilmark{1}, J. W. Park\altaffilmark{1},
D. H. Lee\altaffilmark{2}, W. Han\altaffilmark{2}, U. W.
Nam\altaffilmark{2}, J. H. Park\altaffilmark{2}, \\
J. Edelstein\altaffilmark{3}, E. J. Korpela\altaffilmark{3}, K.
Nishikida\altaffilmark{3}, E. F. van Dishoeck\altaffilmark{4}}

\email{ksryu@satrec.kaist.ac.kr}

\altaffiltext{1}{Korea Advanced Institute of Science and
Technology, 305-701, Daejeon, Korea} \altaffiltext{2}{Korea
Astronomy and Space Science Institute, 305-348, Daejeon, Korea}
\altaffiltext{3}{Space Sciences Laboratory, University of
California, Berkeley, CA 94720} \altaffiltext{4}{Sterrewacht
Leiden, Postbus 9513, 2300 RA LEIDEN, The Netherlands}

\begin{abstract}
The first FUV ($\lambda\lambda$ 1350--1750{\AA}) spectral imaging
observations of the Eridanus superbubble, obtained with the
\emph{SPEAR/FIMS} mission, have revealed distinct fluorescent
emission from molecular hydrogen. In this study, the observed
emission features were compared with those from a photo-dissociation
region model with assumed illuminating stellar fields. The result
showed rather high line ratios of ${\rm I_{1580}/I_{1610}}$, which
may imply the existence of high-temperature molecular clouds in the
region. The H$_2$ fluorescence intensity showed a proportional
correlation with H-$\alpha$ emission, indicating that the
fluorescence and the recombination emission have similar physical
origins.
\end{abstract}


\section{Introduction}

The Eridanus region contains a large nearby interstellar
superbubble. This region was discovered as an extended $15^\circ$
X-ray `hot spot' \citep{williamson1974,naranan_1976}. Subsequently,
the Eridanus Loop was identified as an expanding shell of
H{\footnotesize~I} by \citet{heiles1976} and was extensively imaged
in H-$\alpha$ emission \citep{reynolds1998,boumis2001}, revealing a
large filamentary shell structure. \citet{reynolds_ogden1979} argued
that Barnard's Loop and the Eridanus shell were both created by one
or more supernovae in the Orion OB association and were maintained
by the intense stellar winds and ionizing radiation from those
stars. {\it ROSAT} observations \citep{snowden1995} support the idea
that the huge interstellar cavity is filled with
several-million-degree gas. Surveys of CO \citep{magnani1985,
aoyama2002} show that the cavity is bounded in part by molecular
gas. The CO clouds generally coincide with H{\footnotesize~I}
filaments; however, on a smaller scale, the clouds are located where
H{\footnotesize~I} is weak within the filament. In the FUV spectral
range, early \citep{murthy1993} and improved \citep{kregenow}
spectral observations of diffuse emission from the Eridanus Loop
over limited angular fields show that gas at intermediate
temperatures, T$\sim10^5$K, exists between the hot superbubble
interior and the surrounding material.

We present the first FUV spectral imaging observations of the entire
Eridanus Loop. The data were obtained with the Spectroscopy of
Plasma Emission from Astrophysical Radiation (\emph{SPEAR}), also
known as the Far-ultraviolet Imaging Spectrograph (\emph{FIMS}).
These {\it SPEAR} data show H$_{2}$ fluorescent emission throughout
the Eridanus Loop field and provide new insights into the geometry
and physical conditions of the molecular environment around this
superbubble.

\section{Observations and Data Reduction}

The Eridanus region was observed with \emph{SPEAR}, the main payload
of the Korean satellite, STSAT-1, which was launched on September
27, 2003. \emph{SPEAR} is a dual-channel FUV imaging spectrograph (S
channel $900$--$1150$\AA, L channel $1350$--$1750$\AA,
$\lambda/\Delta\lambda\sim550$, where $\Delta\lambda$ refers to the
half energy width) with a large field of view (S:
$4.0^\circ\times4.6'$, L: $7.5^\circ\times4.3'$, angular
resolution$\sim10'$) optimized for the observation of diffuse
emissions. See \citet{jerrye} for an overview of the instrument,
mission, and data analysis.

The region bounded by $\alpha=2^{\rm h} 30^{\rm m}$ and $5^{\rm h}
30^{\rm m}$, and $\delta=+13^\circ$ and $-23^\circ$, covering most
of the constellations Orion and Eridanus, was observed using more
than 16 ks of exposure taken between 26 December 2003 and 26 January
2004. The observation geometry is shown in Figure \ref{fig1}, along
with an H-$\alpha$ survey map. The H-$\alpha$ features most
prominent around the Eridanus supershell are Barnard's Loop, the I
Orion O-association molecular cloud complex, the H II region
surrounding $\lambda$ Ori centered at $5^{\rm h}32^{\rm m},
+9^{\circ}50^{\rm m}$, and H-$\alpha$ filamentary structures.
Barnard's loop and the $\lambda$ Ori region were excluded from the
{\it SPEAR} observation to avoid excessive count rates. For the
ensuing data analysis, the region was divided, based on the
H-$\alpha$ emission features, into subregions labeled as 1 through 6
(See Figure \ref{fig1}).

We identified bright stars in our observations by examining an FUV
sky intensity map, created by accumulating photons and exposure into
$2'$ sky bins. Stars were identified by matching features, within
$15'$, to the position of UV bright stars in the Tycho-2 catalog
\citep{wright2003}. Photons within $30'$ of the identified stars
were removed from the data set to obtain the residual diffuse
emission and to minimize stellar contamination. After excluding the
contributions of UV bright stars, $3.0\times10^5$ of $8.8\times10^5$
detected photons remained. These photons were accumulated, for each
subregion, into spectra that were binned to 2{\AA}, shown in Figure
\ref{fig2}(a)--(f), along with the noise levels per spectral bin. An
inherent detector background of $\sim0.013$ counts/s/{\AA} was
measured during interleaved shutter-closed periods and was
subtracted from these spectra.

\section{Spectrum Analysis}

The Eridanus 1350{\AA}--1630{\AA} spectra show emission features
that correspond to the H$_{2}$ fluorescence predictions of
\citet{sternberg1989}, notably the prominent bands near 1605{\AA}.
Other prominent features were identified with atomic lines such as
C{\footnotesize~IV} at 1550{\AA} and Si{\footnotesize~II} at
1530{\AA} (See \citet{kregenow} for a discussion of atomic emission
in the Eridanus Loop.). To investigate the physical conditions of
the H$_{2}$, a photo dissociation region (PDR) radiation code
\citep{Black} was used to generate a model FUV spectrum. The
continuum level was taken to be a linear function of wavelength, and
the sensitivity of the fitting while varying the slope and
y-intercept of the continuum was investigated based on the $\chi^2$
minimization method. The fit models for each subregion are shown in
Figure \ref{fig2}.

There are several physical parameters which affect the resultant
spectrum: cloud density, excitation temperature, total hydrogen
molecule column density, and radiation field; we varied only the
excitation temperature in our model with other parameters fixed. For
example, ultraviolet fluorescent emission of H$_{2}$ is produced
following the absorption of photons in the Lyman (B $^1\Sigma_u^+$
-- X $^1\Sigma_g^+$) and Werner (C $^1\Pi_u$ -- X $^1\Sigma_g^+$)
bands. Hence, the stars in the I Orion OB association were
considered as the likely and dominant source of H$_{2}$ emission.
The total radiation field from the most luminous stars of the I
Orion association \citep{reynolds_ogden1979} accounts for the
presumed radiation field scale factor \citep{habing1968} of $\chi
\sim 1.9$, assuming that the distance from the association is
$\sim200$ pc and there is no extinction. \citet{heiles1999}
estimated the electron volume density, $n_e$, of the ionized gas in
the Eridanus superbubble to be 0.8 cm$^{-3}$, which corresponds to
the thermal pressure, $p/k\sim1.5\times10^4$ cm$^{-3}$K, in the
ionized gas. The density of the neutral gas was adjusted to have the
equilibrium thermal pressure, $p/k\sim1.5\times10^4$ cm$^{-3}$K,
according to the excitation temperature, which is used as a
parameter in our model. The fractional abundance of H$_{2}$, defined
as $\langle f \rangle = 2 \langle n({\rm H_2})\rangle/[2\langle
n({\rm H_2})\rangle+\langle n( \mbox{H{\footnotesize~I}})\rangle]$,
was assumed to have an average value of 0.17, as obtained from the
{\it Copernicus} observations \citep{savage1977}. The color
temperature of the incident radiation was assumed to be
$3\times10^4$ K.

With these parameters fixed, the molecular cloud excitation
temperature was varied to investigate the sensitivity of the model
spectrum compared to the observed data. The model spectrum was
scaled to match the average observed intensity over the
1350--1650{\AA} spectral region, where H$_{2}$ emission was apparent
in both the model and the data. The 1356{\AA} airglow line did not
appear in the observed spectrum. The sensitivity of the $\chi^2$
value to the parameters comes from the relative intensity variations
of each vibrational transition band, which are governed by the
distribution of H$_{2}$ among the vibrational excitation states in
the electronic ground state (X $^1\Sigma_g^+$). For example, the
intensity ratio, ${\rm I_{1580}/I_{1610}}$, increases monotonically
from 0.65 to 0.8 as the excitation temperature increases to 1,000 K,
according to our model. Thus, a unique excitation temperature can be
determined if the line ratios are measured.

The results for each subregion, the maximum fluorescent emission
intensity, ${\rm I_{max \, {\rm H}_2}}$, the continuum intensity
$I_{cont}$, and the intensity ratio, ${\rm I_{1580}/I_{1610}}$, are
listed in Table \ref{tbl-1}. The model spectra with T$_{\rm ex} =
50$ K and T$_{\rm ex} = 1,000$ K are plotted  for Region 4 in Figure
\ref{fig2}(g), with the observed spectrum. As can be seen in the
figure, most of the lines in the spectral range are identified as
Lyman bound-bound transitions (BX $v'-v''$), where $v'$ and $v''$
represent the vibrational excitation of a hydrogen molecule at each
respective excitation state. The observed spectrum best fits with a
cloud excitation temperature of 1,000 K, with $\chi_\nu^2$ = 2.2,
though the signal-to-noise ratio of our observation was not
sufficient to discriminate the excitation temperature accurately.

\section{Discussion}

Our detection of H$_2$ fluorescence in Eridanus is consistent with
previous observations in other locations in that fluorescence is
found where H$_2$ is exposed to a strong UV radiation field. The
correlations among the subregions' fluorescent emission intensity,
the H-$\alpha$ emission intensity, and the color excess value are
shown in Figure \ref{fig3}. The H$_2$ fluorescence intensity shows a
strong correlation with both the H-$\alpha$ emission intensity and
the color excess. This suggests that the H-$\alpha$ emission and
hydrogen molecular fluorescence have similar physical origins and
are governed globally by a strong and similar UV radiation field,
likely arising from the I Orion OB association. We expect that the
H$_2$ abundance is  proportional to that of both hydrogen atoms and
dust particles, since dust particles are known as the main birth
place of H$_2$ in the interstellar medium.

\begin{figure}
\epsscale{0.7} \plotone{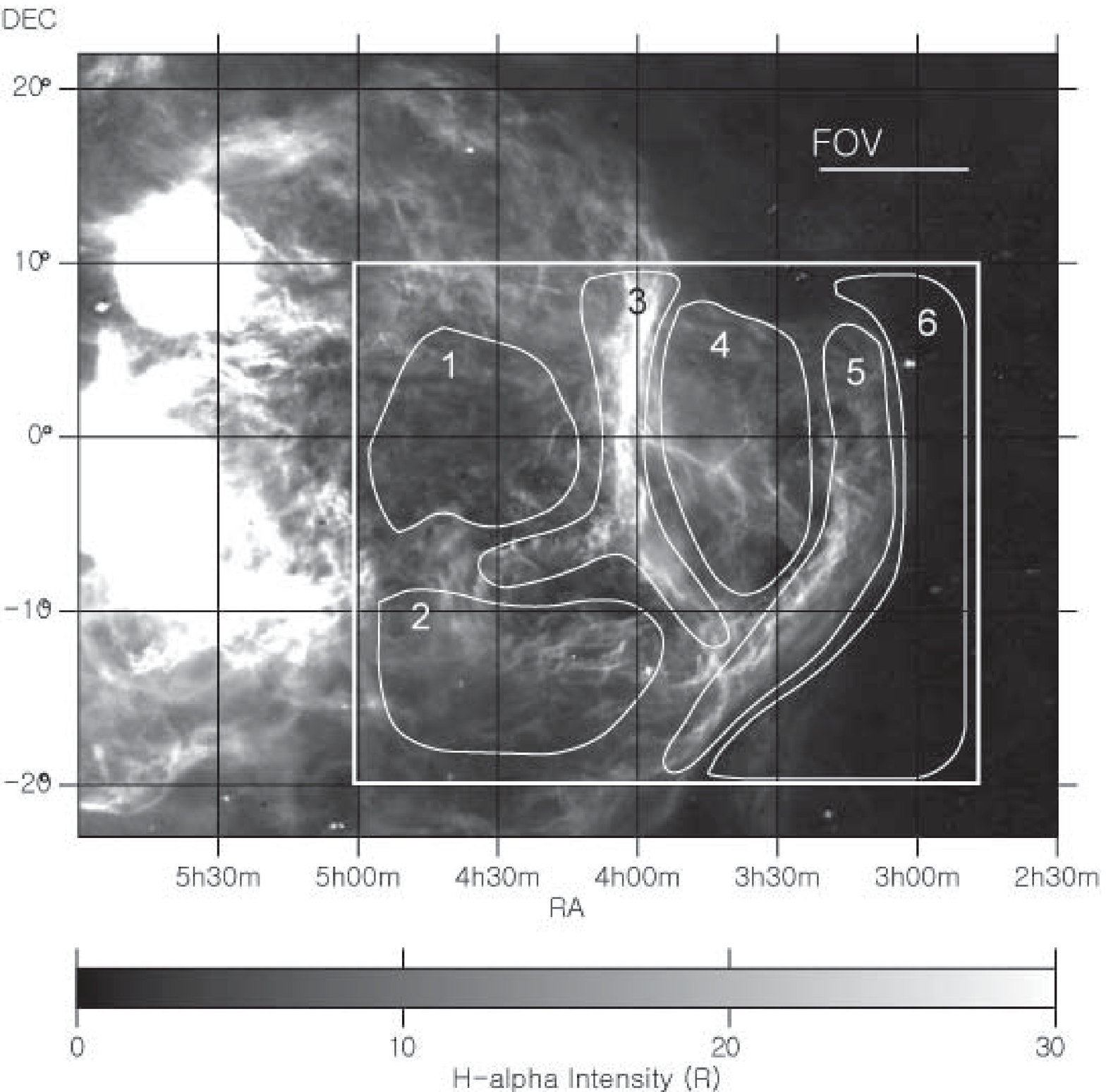} \caption{ Observation geometry of
the Eridanus region marked over an H-$\alpha$ map. The marked
rectangular region was observed with {\it SPEAR}. The observed
region was divided into subregions, labeled 1 through 6, for
spectral analysis and interpretation. The thick solid line, in the
upper right corner, designates the {\it SPEAR} field of view.
\label{fig1}}
\end{figure}

\begin{figure}
\epsscale{0.7} \plotone{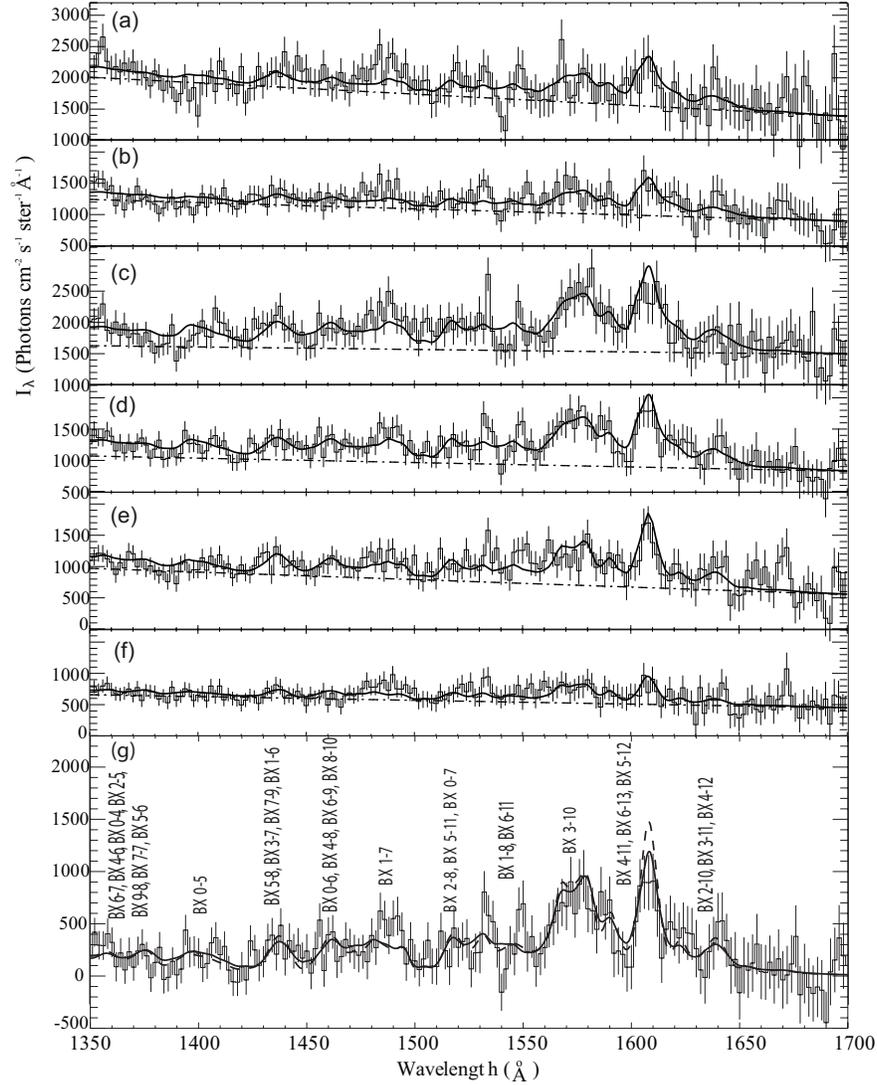} \caption{ {\em SPEAR} FUV diffuse
spectra of Eridanus subregions 1 - 6, shown in panels (a) - (f). The
histogram is the observed spectrum. The thick solid line and the
dashed line are the best-fit model of  H$_2$ fluorescent emission
and continuum intensities, respectively. The noise levels are shown
as vertical bars. (g) The continuum subtracted diffuse spectrum of
Region 4 with different H$_{2}$ model parameters. The dashed line
represents for T$_{\rm ex}=50$ K. The solid line represents for
T$_{\rm ex}=1,000$ K. The legends, marked as BX $v'-v''$, show the
positions of Lyman bound-bound transitions. \label{fig2} }
\end{figure}

\begin{figure}
\epsscale{0.5} \plotone{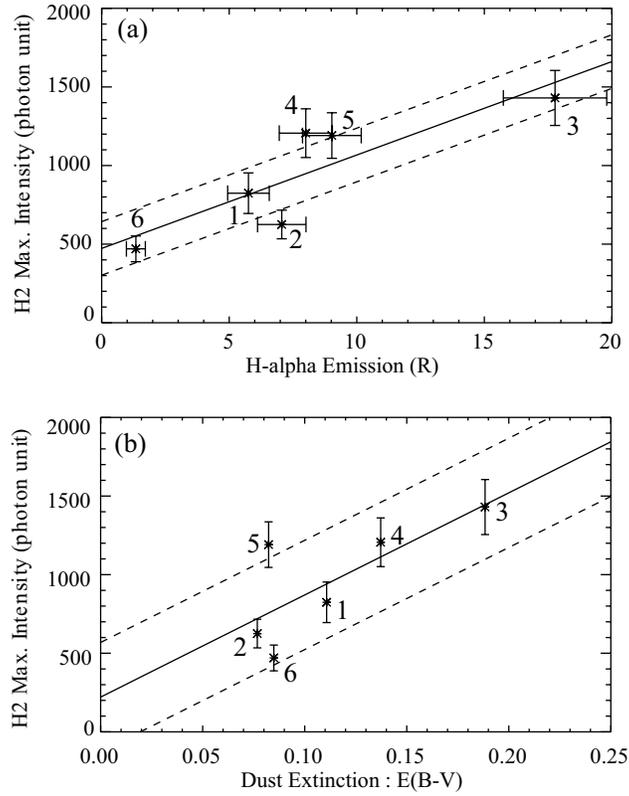} \caption{Comparison of the H$_2$
fluorescent emission intensity (asterisks) for the subregions with
(a) the H-$\alpha$ emission intensity and (b) the color excesses
\citep{finkbeiner2003}. The measurement error ranges are represented
with error bars. The solid lines are the best linear fits, and the
dashed lines show the 1 $\sigma$ uncertainty in $b$, in the linear
fit $y=ax+b$.\label{fig3}}
\end{figure}

The most conspicuous features in the H-$\alpha$ emission map shown
in Figure \ref{fig1}, are the filamentary structures elongated along
$\alpha=4^{\rm h}$(Arc A; Region 3) and $\alpha=3^{\rm h}20^{\rm
m}$(Arc B; Region 5). It appears from Figure \ref{fig3}(a) and (b),
and from Figure \ref{fig1}, that UV photons from the Orion OB
association undergo extinction by the Arc A and Arc B filaments,
since Region 6 shows relatively low H-$\alpha$ and H$_2$
fluorescence, while having a color excess value comparable to those
of Region 1, 2, and 5. The H$_2$ fluorescent emission from Region 5
(Arc B) is relatively high among regions with comparable dust
extinction (Regions 2 and 6), and it is similar to that of Region 4,
even though the dust extinction for Region 4 is higher and the
line-of-sight distance to the I Orion OB FUV sources is likely
smaller. This observation can be explained if the gas in Region 3
(Arc A) is not in the line of sight that extends from the radiation
source (the I Orion OB association) to the gas in Region 5 (Arc B),
i.e., the distances to Arcs A and B are different and the Eridanus
shell structure is open from the I Orion to Arc B. This geometry is
consistent with the considerations of \citet{boumis2001} and
\citet{welsh2005} in that both studies argue that Arcs A and B are
part of a complex of individual shells viewed along the same line of
sight but at different distances (Arc A $>$500pc, Arc B
$\sim$150pc). The observational results do not rule out Reynolds and
Ogden's (1979) description, in that the both Arc A and Arc B can
still be exposed to the radiation from the I Orion OB association
that causes the H$_2$ fluorescent emission.

The FUV continuum intensity does not show a strong correlation with
H-$\alpha$ emission or H$_2$ fluorescent emission intensities (See
Table \ref{tbl-1}). Instead, the continuum intensity seems closely
tied to the geometry of the region. The strong continuum of Region 1
can be attributed to the proximity of Region 1 to the radiation
source. Region 3 is remarkable in that it shows both a strong
continuum intensity and a strong H$_2$ fluorescent emission
intensity.

The observationally derived shell expansion velocity of the Eridanus
region, of between 15 and 23 km s$^{-1}$
\citep{heiles1976,reynolds_ogden1979} is consistent with the
survival of the H$_{2}$ in the Eridanus region. The observed ${\rm
I_{1580}/I_{1610}}$ value suggests the existence of high-temperature
molecular gas in the region. According to \citet{draine1983}, when
an interstellar shock with $v_s = 25$ km/s is running into molecular
gas with a density of $n=1\times10^2$ cm$^{-2}$, the region in which
the gas temperature exceeds 200 K will extend only $2\times10^{16}$
cm, implying that interstellar shocks are not good at producing
large column densities of warm gas. An alternative explanation, for
the heating of molecular gas, might be selective UV excitation by UV
lines radiated by the hot gas in the bubble, or it may be that other
heating mechanisms are able to raise the gas temperature where UV
pumping of the H$_{2}$ is taking place. Further studies are needed
to clarify the contribution and role of UV radiation, supernova
shocks, and other possible heating mechanisms.

In summary, we have presented the first detection of H$_2$
fluorescence in the Eridanus region. The fluorescence emission
features, with a PDR spectrum prediction model, imply the possible
existence of molecular gas with an excitation temperature as high as
1,000 K, which is about an order of magnitude higher than excitation
temperatures generally found for molecular gases in the Galactic
disk \citep{savage1977}. We note that high-excitation temperature
H$_2$ has also been detected in high-velocity shocked gas associated
with the Monoceros Loop \citep{welsh2002} and in a galactic
translucent cloud \citep{snow2000}. The observed spectrum of the
H$_2$ fluorescence in the Eridanus region is different in that a
substantial column density of H$_2$ ($\sim1.0\times10^{20}$
cm$^{-2}$) is required to produce the observed fluorescence. Thus,
most of the H$_2$ undergoing photon pumping must have a high
rotational temperature, if collisional excitation is to account for
the observed fluorescent feature ratios. A more rigorous observation
and model study are expected to reveal the physical conditions of
the molecular gas in the Eridanus supershell.

\acknowledgments \emph{SPEAR/FIMS} is a joint project of KASSI,
KAIST (Korea) and U.C., Berkeley (USA) and is funded by the Korea
MOST and NASA (Grant no. NAG5-5355).

\clearpage

\begin{table}
\begin{center}
\caption{Observational results for regions in Eridanus.
  \label{tbl-1}}
\begin{tabular}{ccccccc}
\tableline\tableline Region &
\multicolumn{1}{c}{I$_{cont.}$\tablenotemark{a}} &
\multicolumn{1}{c}{I$_{\rm max.H_2}$\tablenotemark{b}} &
${\chi_{\nu}^2}$ & ${\rm I_{1580}/I_{1610}}$\tablenotemark{c} &
H-$\alpha$(R) \tablenotemark{d} &
E(B-V) \tablenotemark{e}\\
\tableline 1 & $1590\pm165$ & $825\pm130$ & 2.1 & $0.8\pm0.4$ &
$5.7\pm0.8$ & 0.11 \\
2 & $1065\pm110$ & $625\pm90$ & 2.3 & $0.5\pm0.3$ &$7.0\pm0.9$ & 0.08 \\
3 & $1555\pm160$ & $1430\pm175$ & 2.0 & $1.2\pm0.35$ & $17.7\pm2.0$ & 0.19 \\
4 & $915\pm100$ & $1205\pm155$ & 2.2 & $0.9\pm0.25$ &
$8.0\pm1.0$ & 0.14\\
5 & $720\pm80$ & $1190\pm145$ & 3.8 & $0.85\pm0.25$ &
$9.0\pm1.2$ & 0.08 \\
6 & $520\pm60$  & $470\pm80$ & 2.9 & $1.2\pm0.6$ & $1.3\pm0.4$ &
0.08 \\
\tableline
\end{tabular}
\tablenotetext{a,b}{The intensity of the continuum at 1525{\AA},
I$_{cont}$, and the maximum H$_2$ fluorescent emission from 1600 to
1620{\AA}, I$_{\rm max.H_2}$, in units of
photons/cm$^2$/sr/{\AA}/s.}\tablenotetext{c}{The line ratio, ${\rm
I_{1580}/I_{1610}}$, is estimated by summing 3 spectral bins (6{\AA}
interval), centered at {1580\AA} and {1610\AA}.}
\tablenotetext{d}{The H-$\alpha$ intensity values from
\citet{reynolds1998}} \tablenotetext{e}{The color excess values,
E(B-V) from \citet{finkbeiner2003}.}
\end{center}
\end{table}

\end{document}